\documentstyle[aps,prl,twocolumn,epsf]{revtex}

\begin{document}
\hoffset-0.5cm

\title{Nonperturbative Calculation of the Shear Viscosity
       in Hot $\phi^4$ Theory in Real Time}

\author{Enke Wang$^a$ and Ulrich Heinz$^b$}

\address{$^a$Institute of Particle Physics, Huazhong Normal University,
         Wuhan 430079, China\\
         $^b$Theoretical Physics Division, CERN, CH-1211 Geneva 23, 
         Switzerland}

\date{\today}

\maketitle

\begin{abstract}
Starting from the Kubo formula we calculate the shear viscosity in hot 
$\phi^4$ theory nonperturbatively by resumming ladders with a real-time 
version of the Bethe-Salpeter equation at finite temperature. In the 
weak coupling limit, the generalized Fluctuation-Dissipation Theorem 
is shown to decouple the Bethe-Salpeter equations for the different 
real-time components of the 4-point function. The resulting scalar 
integral equation is identical with the one obtained by Jeon using 
diagrammatic ``cutting rules'' in the Imaginary Time Formalism.  
\end{abstract} 



A systematic field theoretical calculation of the viscosity from the
Kubo formula involving the stress tensor correlation function
\cite{Zubarev} has been the subject of intense theoretical interest
\cite{Hosoya,Jeon1,Wang1,Jeon2}.  First results from a one-loop
calculation \cite{Hosoya,Jeon1,Wang1} of the viscosity in hot $\phi^4$
theory were incomplete because an infinite number of higher order
diagrams in the loop expansion contribute to the lowest order in
powers of the coupling constant \cite{Jeon2}. Their resummation
requires a nonperturbative calculation. Jeon \cite{Jeon2} identified
the dominant class of diagrams contributing to the viscosity at
leading order in the coupling constant and resummed them by using
so-called diagrammatic ``cutting'' rules in the Imaginary Time
Formalism (ITF). In this paper we reproduce his results using an
alternative and, we believe, more economic approach based on a
solution of the Bethe-Salpeter equation in real time using the
Closed Time Path (CTP) formalism \cite{Schwinger,Keldysh,Zhou}.
Our calculation draws heavily on recently derived \cite{Wang3} 
general relations among the different thermal components of 
real-time $n$-point Green functions. 

The CTP formalism has the advantage over the ITF that it is easily
generalized to non-equilibrium situations and that it does not require
analytic continuation of the $n$-point Green functions to real time at
the end of the calculation. It's popularity is, however, decreased by
the doubling of degrees of freedom and the resulting complicated
matrix structure of the Green functions. We begin by reviewing some
important relations among the components of the $n$-point Green
functions in the CTP forma\-lism which help to reduce the complexity of
the real-time calculation before evaluating any Feynman dia\-grams.
In the so-called single-time representation of the CTP formalism the 
$n$-point Green function is defined as
 \begin{equation}
 \label{Ga1an}
   G_{a_1 \dots a_n}(1,\dots,n) \equiv (-i)^{n-1} 
   \langle T_p[\hat\phi_{a_1}(1)\cdots\hat\phi_{a_n}(n)]\rangle \, .
 \end{equation}
Here $\langle\dots\rangle$ denotes the thermal expectation value, 
the numbers $1,\dots,n$ stand for Minkowski space coordinates 
$x_1,\dots,x_n$, $T_p$ represents the time ordering operator 
along the closed time path and corresponds to normal (antichronological) 
time ordering of operators with time arguments on its upper (lower) 
branch, and $a_1,a_2,\dots,a_n=1,2$ indicate on which of the two branches
the fields are located. (For details see, e.g., \cite{Zhou,Wang3}.) 
Using the KMS condition \cite{Kubo} one derives in momentum space 
\cite{Zhou}
 \begin{eqnarray}
 \label{G*}
    &&G^*_{a_1 a_2\dots a_n}(k_1, k_2,\dots, k_n)  =
  \nonumber\\
    &&(-1)^{n-1} 
      G_{{\bar a}_1 {\bar a}_2\dots {\bar a}_n}(k_1, k_2,\dots, k_n)
      \prod\limits_{\{i|a_i=2\}} \!\!\! e^{\beta k^0_i}\, ,
  \end{eqnarray} 
where $k_1{+}k_2{+}\dots{+}k_n=0$, the star denotes complex conjugation, 
$\beta$ is the inverse temperature, and ${\bar a}_i=2,1$ for $a_i=1,2$, 
respectively. The ``physical'' or $r/a$ representation is defined by 
setting \cite{Zhou}
 \begin{equation}
 \label{ar} 
  \hat\phi_a(x) = \hat\phi_1(x){-}\hat\phi_2(x) ,\ \  
  \hat\phi_r(x)={\hat\phi_1(x){+}\hat\phi_2(x)\over 2} ,
 \end{equation}
and writing the $n$-point Green function as
 \begin{equation}
 \label{Gaf1afn}
  G_{\alpha_1\dots\alpha_n}(1,\dots,n) \equiv 
  {2^{n_r-1}\over i^{n-1}} \langle 
  T_p[\hat\phi_{\alpha_1}(1)\cdots\hat\phi_{\alpha_n}(n)]\rangle \, .
 \end{equation}
Here $\alpha_1,\dots,\alpha_n=a,r$, and $n_r$ is the number 
of $r$ indices among $(\alpha_1,\alpha_2,\dots,\alpha_n)$. 
One can show \cite{Zhou} that
 \begin{equation}
 \label{Gaaa}
   G_{aa\dots a}(1,\dots,n)=0\, ,
 \end{equation}
that the functions with only one $r$ index, $G_{ra\dots a}$,
$G_{ara\dots a}$, \dots, $G_{a\dots ar}$, are the fully retarded
$n$-point Green functions defined in \cite{Lehmann}, and that (except
for $G_{rr\dots r}$) all other components
$G_{\alpha_1\dots\alpha_n}(1,\dots,n)$ involve both retarded and
advanced relations among their $n$ time arguments. The simplest
example is the 2-point function:
 \begin{eqnarray}
 \label{GrGaa}
   &&\Delta_{aa}(k) = 0,\ 
     \Delta_{ra}(k) = \Delta^{\rm ret}(k),\  
     \Delta_{ar}(k) = \Delta^{\rm adv}(k) , 
 \\
 \label{GrGab}
   &&\Delta_{rr}(k) = \bigl( 1{+}2n(k^0) \bigr)
     \bigl( \Delta_{ra}(k)-\Delta_{ar}(k) \bigr)\, .
 \end{eqnarray} 
Eq.~(\ref{GrGab}) is the fluctuation-dissipation theorem \cite{Callen}, 
with $n(k^0)=1/(e^{\beta k^0}-1)$. The retarded and advanced 2-point 
functions satisfy
 \begin{eqnarray}
   &&\Delta^*_{ra}(k) = \Delta_{ar}(k) , 
 \label{raa}\\
   &&\Delta_{ra}(-k) = \Delta_{ar}(k), \ \ 
     \Delta_{rr}(-k) = \Delta_{rr}(k) .  
 \label{rab}
 \end{eqnarray} 
The the single-time and $r/a$ representations of the $n$-point Green 
functions are related by \cite{Zhou}
 \begin{eqnarray}
 \label{Gtrans}
   G_{a_1 \dots a_n}(1,\dots,n) &=&
   2^{1-{n\over 2}} G_{\alpha_1\dots\alpha_n}(1,\dots,n)
 \nonumber\\
   &&\times\,  Q_{\alpha_1 a_1} \cdots Q_{\alpha_n a_n}\, ,
 \end{eqnarray}
where repeated indices are summed over and 
 \begin{equation}
 \label{Q}
    Q_{a 1}=-Q_{a 2}=Q_{r 1}=Q_{r 2}={1\over \sqrt{2}}
 \end{equation}
are the four elements of the orthogonal Keldysh transformation for 2-point
functions\cite{Kubo}. $n$-point Green functions involving only a single
field $\phi$ are symmetric under particle exchange:
 \begin{eqnarray}
 \label{Gsym}
   &&G_{\dots\alpha_i\dots\alpha_j\dots}(\dots,k_i,\dots,k_j,\dots)=
 \nonumber\\   
   &&G_{\dots\alpha_j\dots\alpha_i\dots}(\dots,k_j,\dots,k_i,\dots)
 \end{eqnarray}

In the following we study massless $\phi^4$ theory in the weak coupling
limit:
 \begin{equation}
 \label{lag0}
    {\cal L}_0 = {1\over 2} (\partial_{\mu}\phi)^2 + 
                 {\lambda \over 4!} \phi^4, \qquad \lambda\ll 1.
 \end{equation} 
The Kubo formula for the shear viscosity \cite{Hosoya,Jeon1}
 \begin{equation}
 \label{eta1}
   \eta = {\beta \over 20}\lim_{p^0,\bbox{p} \to 0}\int d^4x\, 
   e^{ip{\cdot}x} \langle\pi_{lm}(x)\pi^{lm}(0)\rangle 
 \end{equation}
involves the correlation function of the traceless pressure tensor 
$\pi_{lm}$ which in the thermal rest frame reads 
 \begin{equation}
 \label{pilm}
   \pi_{lm}(x) = \left(\delta_{li}\delta_{mj}-
   {\textstyle{1\over 3}}\delta_{lm}\delta_{ij}\right)\,
   \partial_i\phi\,\partial_j\phi\,
 \end{equation}
with $i,j,l,m=1,2,3$. From the definitions (\ref{Ga1an},\ref{eta1}) 
one sees that $\eta$ is related to the 12-component of the 2-point Green 
function $\Delta_{\pi\pi}$ of the composite field $\pi_{lm}$ in the 
single-time representation:
 \begin{equation}
 \label{eta2}
   \eta = {{i\beta}\over 20}\lim_{p^0,\bbox{p} \to 0}
          \Delta_{\pi\pi}^{12}(p)
   = -{{\beta}\over 20}\lim_{p^0,\bbox{p} \to 0}
         {\rm Im\,} \Delta_{\pi\pi}^{12}(p)\,.
 \end{equation}
In the second equation we used that the (12)-component of any 2-point
function is purely imaginary as follows from 
Eqs.\,(\ref{GrGab})-(\ref{Gtrans}). In (\ref{pilm}) each composite 
field $\pi_{lm}$ is composed of two single particle fields $\phi$. 
Substituting (\ref{pilm}) into (\ref{eta1}) one can rewrite 
(\ref{eta2}) as
 \begin{eqnarray}
 \label{eta3}
   \eta &=& {{\beta}\over 5}\lim_{p^0,\bbox{p} \to 0}
          \int{d^4k\over (2\pi)^4}\int{d^4q\over (2\pi)^4} 
          J_{lm}(-k, p{+}k)
 \nonumber\\
 &&\times\, {\rm Im\,}G_{1122}(-k, p{+}k, q, -p{-}q) J^{lm}(q, -p{-}q)
 \end{eqnarray}
where $G_{1122}$ denotes the $(1122)$-component of the 4-point Green 
function for the $\phi$ field, $J_{lm}(p,q)$ joins two $\phi$ 
propagators to a point (see Fig.\,1): 
 \begin{equation}
 \label{join}
   J_{lm}(p,q) = p_l q_m-{\textstyle{1\over 3}}\delta_{lm}\,
   \bbox{p}{\cdot}\bbox{q}\,,
 \end{equation}
and a symmetry factor 4 accounts for the different possibilities of 
doing so.

Jeon showed \cite{Jeon2} that in $\phi^4$ theory at leading order of
the coupling constant ($\sim{1\over \lambda^2}$) all planar ladder
diagrams (see Fig.~1) contribute to the shear viscosity. On the other
hand, contributions from crossed diagrams can be ignored. In the CTP
formalism the infinite number of ladder diagrams can be resummed by
using the Bethe-Salpeter (BS) equation. Fig.~2 illustrates the
correspon\-ding BS integral equation for the 4-point Green function,
with ladders consisting of rungs formed by a one-loop dia\-gram
connecting two propagators. In general this BS integral equation will
couple the different components of the 4-point Green function in the
CPT formalism with each other. The key technical issue is therefore
whether and how these equations can be decoupled. In the following we
will show that in the weak coupling limit this is indeed possible, and
that the most convenient basis for doing so is the $r/a$
representation of the CTP formalism.  The previously derived
generalized fluctuation-dissipation theorem (FDT) \cite{Wang3} plays
an important role in the decoupling procedure.

 \begin{figure}
 \epsfxsize 80mm \epsfbox{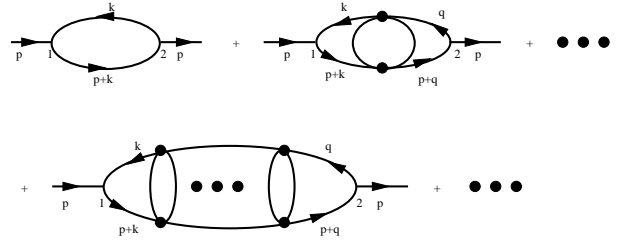}
 \vskip 0.4cm
 \caption{The planar ladder diagrams contributing to the shear viscosity
           at leading order.
      \label{F1}}
 \end{figure}

 \begin{figure}
  \epsfxsize 80mm \epsfbox{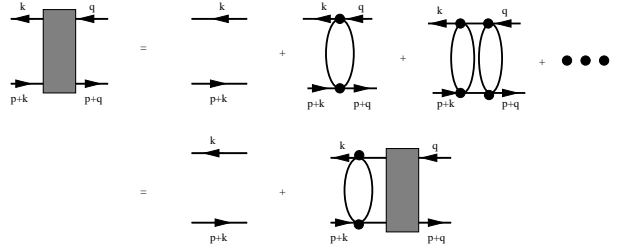}
  \vskip 0.4cm
  \caption{Bethe-Salpeter equation for the 4-point Green function.
      \label{F2}}
 \end{figure}

From Eq.\,(\ref{G*}) we deduce
 \begin{eqnarray}
 \label{G1122}
   &&G_{1122}(-k,p{+}k,q,-p{-}q)= -n(p^0)
 \\
   &&\times [G_{1122}(-k,p{+}k,q,-p{-}q)
            +G^*_{2211}(-k,p{+}k,q,-p{-}q)] 
 \nonumber
 \end{eqnarray}
so that Eq.\,(\ref{eta3}) can be rewritten as
 \begin{eqnarray}
 \label{eta4}
   \eta &=& -{{\beta}\over 5}\lim_{p^0,\bbox{p} \to 0}n(p^0)
            \int_{k,q} J_{lm}(-k,p{+}k) J^{lm}(q, -p{-}q)
 \nonumber\\
   && \ \times {\rm Im\,} [G_{1122}{+}G^*_{2211}](-k,p{+}k,q,-p{-}q)].
 \end{eqnarray}
In the physical representation we have (cf. Eq.\,(\ref{Gtrans}))
 \begin{eqnarray}
 \label{G1122+}
   &&{\rm Im\,}[G_{1122}+G^*_{2211}] =
    -{\textstyle{1\over 4}}{\rm Im\,}[G_{rrra}{+}G_{aaar}
 \\
   &&\quad 
   {+}G_{rrar}{+}G_{aara}{-}G_{rarr}{-}G_{araa}{-}G_{arrr}{-}G_{raaa}]\, .  
 \nonumber
 \end{eqnarray}
From the generalized FDT \cite{Wang3} we know that only 7 of the 16 
components of $G_{\alpha_1\alpha_2\alpha_3\alpha_4}$ are independent;
choosing them as the 4 fully retarded functions plus $G_{arra}$, 
$G_{arar}$ and $G_{aarr}$ and using the relations derived in Sec. III\,C 
of Ref.\,\cite{Wang3} we find
 \begin{eqnarray}
 \label{imaginary}
  && {\rm Im\,}\bigl(G_{1122}{+}G^*_{2211}\bigr)=
     {\rm Im\,}\bigl(a G_{raaa} + b G_{araa} + c G_{aara}
 \nonumber\\  
  &&\qquad + d G_{aaar} + e G_{arra} + f G_{arar}+ g G_{aarr}\bigr)\, ,
 \end{eqnarray}
where the coefficients $a,b,\dots,g$ are sums and products of thermal 
distribution functions. In the limit $p^0{\to}0$ the latter are given by
 \begin{mathletters}
 \label{coeff}
 \begin{eqnarray}
  a &=& \beta p^0\, N_k (N^2_q{-}1) 
        + {\cal O}\bigl((\beta p^0)^2\bigl)\, ,
 \label{coeff1}\\
  b &=& {(\beta p^0)^2\over 4} (N^2_k{-}1)(N^2_q{-}1) 
        + {\cal O}\bigl((\beta p^0)^3\bigl)\, ,
 \label{coeff2}\\
  c &=& {\beta p^0\over 2} \left[ N_k (N^2_q{-}1)+ N_q (N^2_k{-}1)\right] 
        + {\cal O}\bigl((\beta p^0)^2\bigl)\, ,
 \label{coeff3}\\
  d &=& {\beta p^0\over 2} \left[ N_k (N^2_q{-}1)- N_q (N^2_k{-}1)\right]
        + {\cal O}\bigl((\beta p^0)^2\bigl)\, ,
 \label{coeff4}\\
  e &=& {\beta p^0\over 2} \left[ (N^2_k{-}1)+ (N^2_q{-}1)\right]  
        + {\cal O}\bigl((\beta p^0)^2\bigl)\, ,
 \label{coeff5}\\
  f &=& {\beta p^0\over 2} \left[ (N^2_k{-}1)+ (N^2_q{-}1)\right]
        + {\cal O}\bigl((\beta p^0)^2\bigl)\, ,
 \label{coeff6}\\
  g &=& -\beta p^0 (N^2_k{-}1)      
        + {\cal O}\bigl((\beta p^0)^2\bigl)\, ,
 \label{coeff7}
 \end{eqnarray} 
 \end{mathletters}
where $N_k=1+2 n(k^0)$. With $\lim_{p^0\to 0} \beta p^0\, n(p^0)=1$ 
and using the symmetry of the integrations over $k$ and $q$ in 
Eq.\,(\ref{eta4}) as well as the symmetry relation (\ref{Gsym}),
the shear viscosity (\ref{eta4}) is found as 
 \begin{eqnarray}
 \label{eta5}
   \eta &=& {\beta\over 5}\int{d^4k\over (2\pi)^4}
            n(k^0)\bigl(1{+}n(k^0)\bigr)I_{\pi,lm}(k)
 \nonumber\\
   &&\times\int{d^4q\over (2\pi)^4} 
     {\rm Im\, }{\bar G}(-k,k,q,-q)I_\pi^{lm}(q)\, ,
 \end{eqnarray}
where $I_{\pi,lm}(k)\equiv -J_{lm}(-k,k)$ as in \cite{Jeon2} and 
 \begin{equation}
 \label{Gbar}
   {\bar G} = 2\, G_{arra} - G_{aarr}\, .
 \end{equation}

We now set up the BS integral equation for $\bar G$ according to 
Fig.\,2. The Feynman rules of the $r/a$ representation \cite{Wang3} 
give
 \begin{eqnarray}
 \label{Galpha}
   &&i^3 G_{\alpha_1\alpha_2\alpha_3\alpha_4}(-k,k,q,-q)=
 \\
   && \left[i\Delta_{\alpha_1\alpha_3}(-k)\right]
      \left[i\Delta_{\alpha_2\alpha_4}(k)\right]
      (2\pi)^4\delta^4(k{-}q)+
 \nonumber\\
    &&{1\over 2}\left[i\Delta_{\alpha_1\beta_1}(-k)\right]
    \left[i\Delta_{\alpha_2\gamma_1}(k)\right]
    (-i\lambda_{\beta_1\beta_2\beta_3\beta_4})
    (-i\lambda_{\gamma_1\gamma_2\gamma_3\gamma_4})
 \nonumber\\
    && \ \times
       \int{d^4s\over (2\pi)^4}\,{d^4l\over (2\pi)^4}
       \left[i\Delta_{\gamma_2\beta_2}(s)\right]
       \left[i\Delta_{\beta_3\gamma_3}(s+l-k)\right]
 \nonumber\\
    && \qquad \times 
       \left[i^3 G_{\beta_4\gamma_4\alpha_3\alpha_4}(-l,l,q,-q)\right].
 \nonumber
 \end{eqnarray}
Here the factor ${1\over 2}$ on the r.h.s. is the symmetry factor 
associated with the bubble connecting the two lines, and in the 
$r/a$ representation the bare 4-point vertex is given by \cite{Wang3} 
 \begin{equation}
 \label{coupling}
  \lambda_{\alpha_1\alpha_2\alpha_3\alpha_4}=
  {\lambda\over 4}\left[1-(-1)^{n_a}
   \right]\, .
 \end{equation}
$n_a$ is the number of $a$ indices among $(\alpha_1,\alpha_2,
\alpha_3,\alpha_4)$. 

In general, the full retarded 2-point Green function $\Delta_{ra}(k)$ 
can be expressed as
 \begin{equation}
 \label{fullret}
  \Delta_{ra}(k)={1\over {k^2+{\rm Re\, }\Sigma(k)
           +i\,{\rm Im\, }\Sigma(k)}}\, .
 \end{equation}
The advanced Green function $\Delta_{ar}(k)$ is given by the complex 
conjugate expression (see (\ref{raa})). For massless $\phi^4$ theory 
Re\,$\Sigma(k) = -\lambda T^2/24 + {\cal O}(\lambda^2)$ whereas 
Im\,$\Sigma(k) \sim {\cal O}(\lambda^2)$\cite{Parwani,Wang2}. For weak 
coupling the propagators thus have quasiparticle poles at Re\,$E_k =
\pm\sqrt{\bbox{k}^2+\lambda T^2/24}$ with small imaginary parts, and
in the limit $\lambda \to 0$ the product $\Delta_{ra}(k)\Delta_{ar}(k)$ 
develops a pinch singularity at $E_k$ due to the converging pair of 
poles in the upper and lower half of the complex energy plane. Since 
the products $\Delta_{ra}(k)\Delta_{ra}(k)$ and $\Delta_{ar}(k)
 \Delta_{ar}(k)$ have no pinch singularities, the corresponding 
contributions to the shear viscosity can be neglected for 
$\lambda\ll 1$. 

We now evaluate the BS equation for $\bar G$ in this appro\-ximation. 
Using (\ref{GrGaa})-(\ref{rab}) in Eq.\,(\ref{Galpha}) one obtains 
for the first term on the r.h.s. of (\ref{Gbar}) 
 \begin{eqnarray}
 \label{Garra}
   &&G_{arra}(-k,k,q,-q) = {\lambda^2\over 4}
     N(k^0)\Delta_{ra}(k)\Delta_{ar}(k)
 \\
    &&\times \int{d^4s\over (2\pi)^4}\,{d^4l\over (2\pi)^4}
    \Delta_{rr}(s)\Delta_{rr}(s{+}l{-}k)
    G_{aara}(-l,l,q,-q)\, .
 \nonumber
 \end{eqnarray}
Obviously this doesn't decouple: $G_{arra}$ couples to $G_{aara}$. 
However, changing the integration variables $s\to {-}s$ and 
$l\to {-}l$ and using the relations (\ref{rab}) and (\ref{Gsym}) 
together with $N(-k^0)=-N(k^0)$ one finds that $G_{arra}(-k,k,q,-q)$ 
is an odd function of $k$. The corresponding contribution to the shear 
viscosity (\ref{eta5}) thus vanishes by symmetric integration. 

For the second term in (\ref{Gbar}) we obtain in the same approximation 
from Eq.\,(\ref{Galpha})
 \begin{eqnarray}
 \label{Gaarr}
    &&G_{aarr}(-k,k,q,-q) =  -\Delta_{ra}(k)\Delta_{ar}(k)
    \Bigl\{ i(2\pi)^4\delta^4(k{-}q)
 \nonumber\\     
    &&+{\lambda^2\over 8} \int{d^4s\over (2\pi)^4}\,{d^4l\over (2\pi)^4}
    G_{aarr}(-l,l,q,-q)
 \nonumber\\
    &&\times\Bigl[ \Delta_{ra}(s)\Delta_{ar}(s{+}l{-}k)
                 +\Delta_{ar}(s)\Delta_{ra}(s{+}l{-}k)
 \nonumber\\
    &&\quad +\Delta_{rr}(s)\Delta_{rr}(s{+}l{-}k)
 \nonumber\\
    &&\quad +N(l^0)\bigl( \left[\Delta_{ra}(s)-\Delta_{ar}(s)\right]
                  \Delta_{rr}(s{+}l{-}k)
 \nonumber\\
    &&\quad +\Delta_{rr}(s)\left[\Delta_{ar}(s{+}l{-}k)
            -\Delta_{ra}(s{+}l{-}k)\right]\bigr)
    \Bigr]\Bigr\}
 \end{eqnarray}
Already here one sees that the BS equation decouples: the integral
equation involves only the single component $G_{aarr}$ of the 4-point 
function. To make further progress it is convenient to introduce a 
function $M$ which is obtained by truncating two of the external legs 
of the 4-point Green function $G$:
 \begin{eqnarray}
 \label{M}
  &&G_{\alpha_1\alpha_2\alpha_3\alpha_4}(-k,k,q,-q) =
 \\
  && \left[i\Delta_{\alpha_1\beta_1}(-k)\right]
    \left[i\Delta_{\alpha_2\beta_2}(k)\right]
    M_{\beta_1\beta_2\alpha_3\alpha_4}(-k,k,q,-q)\, .
 \nonumber
 \end{eqnarray}
We also introduce the 2-point spectral density
 \begin{equation}
 \label{density}
    \rho(k)=i\left[\Delta_{ra}(k)-\Delta_{ar}(k)\right]
 \end{equation}
and derive from Eq.\,(\ref{raa}) 
 \begin{equation}
 \label{pinch}
   \Delta_{ra}(k)\Delta_{ar}(k)={\rho(k)\over {2{\rm Im\, }\Sigma(k)}}.
 \end{equation}
In the pinching pole approximation the first two terms in the square 
brackets in Eq.\,(\ref{Gaarr}) can be approximated as
 \begin{eqnarray}
 \label{approx}
   &&\Delta_{ra}(s)\Delta_{ar}(s{+}l{-}k)
    +\Delta_{ar}(s)\Delta_{ra}(s{+}l{-}k)
 \nonumber\\
   && \approx [\Delta_{ra}(s)-\Delta_{ar}(s)]
     [\Delta_{ar}(s{+}l{-}k)-\Delta_{ra}(s{+}l{-}k)]
 \nonumber\\
   && = \rho(s)\rho(s{+}k{-}l)\, .
 \end{eqnarray}
With these ingredients, and using Eq.\,(\ref{GrGab}), Eq.\,(\ref{Gaarr}) 
can then be simplified as
 \begin{eqnarray}
 \label{Mrrrr}
    &&{\rm Im\, }M_{rrrr}(-k,k,q,-q)=(2\pi)^4\delta^4(k{-}q)+
 \nonumber\\ 
    &&{\lambda^2\over 4}\int{d^4s\over (2\pi)^4}\,{d^4l\over (2\pi)^4}\,
    {{\rho(l) \, \rho(s) \, \rho(s{+}k{-}l)}
     \over {{\rm Im\, }\Sigma(l)}}\times
 \\
    &&{{[1{+}n(l^0)][1{+}n(s^0{+}k^0{-}l^0)]n(s^0)}\over {1+n(k^0)}}
    {\rm Im\, }M_{rrrr}(-l,l,q,-q)\, .
 \nonumber
 \end{eqnarray}
Note that the integral equation remains decoupled when expressed 
through the truncated 4-point function $M$.

For the shear viscosity one thus gets
 \begin{eqnarray}
 \label{eta6}
   \eta &=& {\beta\over 10}\int{d^4k\over (2\pi)^4}\,
          n(k^0)[1{+}n(k^0)]I_{\pi}(k)
          {\rho(k)\over {\rm Im\, }\Sigma(k)}
 \nonumber\\
   &&\times \int{d^4q\over (2\pi)^4}I_{\pi}(q) 
          {\rm Im\, }M_{rrrr}(-k,k,q,-q)\, .
 \end{eqnarray}
This can be brought into the form given by Jeon \cite{Jeon2} by 
defining
 \begin{equation}
 \label{Dpi}
   D_{\pi}(k) = \int{d^4q\over (2\pi)^4}\,I_{\pi}(q)\, 
                {\rm Im\, }M_{rrrr}(-k,k,q,-q)\, .
 \end{equation}
Inserting Eq.~(\ref{Mrrrr}) into this definition, the BS equation 
derived here is found to coincide with the results obtained by Jeon 
in Eqs.\,(4.16)--(4.21) of Ref.\,\cite{Jeon2}.

Let us summarize: We have shown that a nonperturbative calculation of 
the shear viscosity in hot $\phi^4$ theo\-ry can be performed in the 
real-time (CTP) formalism by solving a BS equation. In the weak 
coupling limit the BS integral equation decouples and involves only
a single component of the real-time thermal 4-point function. Our
results agree with those of Jeon \cite{Jeon2}, but our derivation is
much more compact. This was made possible by using the generalized 
fluctuation-dissipation theo\-rem which relates the different components
of real-time thermal $n$-point functions. 

\acknowledgments
This work was started when E.~W. worked as a Humboldt Fellow at the 
Institut f\"ur Theoretische Physik, Universit\"at Regensburg. He thanks 
the Alexander von Humboldt Foundation for their support and also
acknowledges partial support by the National Natural Science Foundation 
of China (NSFC) and the Science Research Foundation of Hubei Province in 
China. The work of U.~H. was supported in part by DFG, BMBF and GSI.
The authors thank the Institute for Nuclear Theory at the University 
of Washington and the organizers of the workshop INT-99-3 for their  
hospitality and the Department of Energy for partial support during 
the completion of this work.


\end{document}